# A Multi-Faceted Approach to Scrutinizing the Reliability of a Measure of STEM Teacher Strategic Knowledge

Robert M. Talbot III, PhD[1]

**Abstract**
Score reliability is necessary for establishing a validity argument for an instrument, and is therefore highly important to investigate. Depending on the proposed instrument use and score interpretations, differing degrees of precision in measurement or reliability are required. Researchers sometimes fail to take a critical stance when investigating this important measurement property, and default to accepted values of commonly known measures. This study takes a multi-faceted approach to scrutinizing score reliability from a measure of STEM teachers' strategic knowledge using rater agreement, classical test theory conceptions of reliability, and Generalizability Theory. This detailed examination provides insight about where the greatest gains in score reliability can be realized, given the design of the instrument and the context of measurement.

**Introduction**

Score reliability is foundational to much of the work we do in science education research. Researchers often speak of needing instruments that are "valid and reliable", but that phraseology does not fully describe what we need in a measurement instrument. For example, "reliability" is a property of the scores resulting from responses to the instrument, not a property of the instrument itself. The term "validity" should be used to denote the "degree to which evidence and theory support the interpretation of test scores entailed by the proposed test uses" (American Educational Research Association, American Psychological Association, & National Council on Measurement in Education, 2014). Score reliability can be considered a necessary condition for validity, and is therefore highly important to investigate. Depending on the proposed instrument use and score interpretations, differing degrees of precision in measurement or reliability can be required for making a validity argument. While many procedures and estimation methods are available for researchers to use in characterizing score reliability, we often fail to take a critical stance when investigating this important measurement property, and default to accepted values of commonly known measures.

In this study, I take a critical approach to scrutinizing score reliability from a scenario-based measurement of strategic knowledge (SK). This instrument was developed to evaluate the effect of a university level teaching experience on novice STEM teachers' SK. By taking this deliberate approach to examining reliability, instrument potential and limitations are uncovered which help to establish validity.

---

[1] University of Colorado Denver, School of Education and Human Development, robert.talbot@ucdenver.edu

**Measurement Context: Strategic Knowledge Construct**

The strategic knowledge construct is composed of two dimensions: the Flexible Application (FA) of instructional approaches, and the use of Student Centered Instruction (SCI), hence the instrument is named the FASCI. Strategic knowledge is closely related to pedagogical content knowledge (PCK, Shulman, 1986), particularly with respect to two characteristics: Part of pedagogical content knowledge involves (a) teachers' representations of subject matter and strategies for teaching the subject matter (related to the FA dimension), and (b) teachers' knowledge of students' understanding of the subject matter (related to the SCI dimension),. Although the construct of strategic knowledge is related to pedagogical content knowledge, the FASCI is not intended to measure PCK.

The latent variable underlying each of the dimensions of strategic knowledge (FA or SCI) can be described using a construct map (Wilson, 2005). A construct map describes the qualitatively distinct levels that are hypothesized to exist on the continuum of each latent variable. The construct maps for each of the FASCI dimensions are shown in Figures 1 (the FA construct map) and 2 (the SCI construct map).

**Flexible Application.**

The Flexible Application (FA) dimension of strategic knowledge describes the strategic repertoire that a teacher possesses and how (at the highest level) she makes decisions based on relevant contextual factors to employ these instructional strategies. A teacher who is at an expert level on the FA dimension (level "2") is an adaptive expert (Hatano & Inagaki, 1986) in that she has a large "adaptational repertoire" of teaching strategies like the experts in the study by Clermont, Borko, and Krajcik (Clermont, Borko, & Krajcik, 1994). These strategies are based in the teachers' knowledge of representations of the subject matter, a characteristic of PCK and an essential component of the pedagogical reasoning process (Shulman, 1987). A teacher at this highest level is able to consider relevant constraints within their area, which is consistent with Chi's view of expertise (Chi, 2006). At a slightly less sophisticated level on the FA dimension (level "1"), a teacher has a repertoire of instructional strategies but chooses a strategy without consideration of relevant contextual factors (such as student understanding). The teacher can adapt a teaching strategy as needed, but does not necessarily have a strong rationale for this modification. At the lowest level on the FA dimension (level "0"), a teacher has a very limited repertoire of strategies to choose from and once she has decided on a strategic approach for a given context, she does not adapt or change it. Though this dimension is conceptualized as being continuous in nature, instrument developers and response raters defined these three qualitatively distinct levels of the construct.

| Level | Respondent Characteristics |
|---|---|
| 2 | ● The teacher has repertoire of strategies that can be used to facilitate student learning within a given class session.<br>● If the teaching strategy comprised of these acts is not producing the desired result, sometimes it can be modified.<br>● The teacher recognizes that the choice of a class activity and associated teaching strategy will depend upon variables specific to the classroom context. |
| 1 | ● The teacher has a repertoire of strategies that can be used to facilitate student learning within a given class session.<br>● If an activity based on a particular teaching strategy is not producing the desired result, the activity can be modified by selecting a different strategy. |
| 0 | ● The teacher has a limited repertoire of strategies.<br>● Once a particular activity has been selected for a class session, it is not easily modified with a different strategy. |

*Figure 1.* Construct map for the Flexible Application (FA) dimension

**Student Centered Instruction.**

The student centered instruction (SCI) dimension describes how a teacher views an activity as an opportunity for students active engagement so that students' ideas are elicited and articulated. Eliciting and building on student prior knowledge is central to teaching for understanding (Bransford, Brown, & Cocking, 2000; Collins, Greeno, Resnick, Berliner, & Calfee, 1992; Fosnot, 1996). At the expert level (level "2"), a teacher conceives of the situation as an opportunity for interaction between herself and the students, or between the students and each other. She also articulates a rationale for why she conceives of the situation in this way. A teacher at the middle level (level "1") views the learning activity as a situation in which she and/or the students are interacting with each other. This level of thinking was identified in a study by Peterson and Treagust (Peterson & Treagust, 1995). In this study, as teachers engaged in Shulman's pedagogical reasoning process, they became more student-centered. At a novice level on the SCI dimension (level "0") the teacher views the learning activity as a non-interactive place where she presents the material to her students without any adaptation or tailoring of the material and representations to her students' needs. Again, although the SCI dimension is conceptualized as being continuous in nature, instrument developers and response raters defined these three qualitatively distinct levels of the construct.

| Level | Respondent Characteristics |
|---|---|
| 2 | ● Discussion of interactive teaching which would be observable to the teacher or to an outside "other."<br>● Discussion of a rationale for why they see this as an opportunity for interactive teaching and learning |
| 1 | ● Discussion of interactive teaching which would be observable to the teacher or to an outside "other." |
| 0 | ● No discussion of interactive teaching<br>● Teacher primarily views classroom activities as ways to help students make sense of new ideas. Information goes from teacher to student. |

*Figure 2*. Construct map for the Student-Centered Instruction (SCI) dimension

**Instrument Design.**

The FASCI instrument includes open-ended scenario-based items. These items are broadly contextualized in terms of the content being taught. In other words, no specific STEM content domain is specified in the scenarios. An example FASCI item is shown in Figure 3. The five-item version of the FASCI used in this study can be seen in Appendix A.

---

**Example FASCI item**
**For the questions and scenarios that follow, please assume that you are teaching a high school course in physics, chemistry, biology, earth science or math to a class of 25-30 students.**

Students are working in groups of four to discuss a conceptual question you provided them at the beginning of class.
  a.) How might this activity facilitate student learning?

As the activity proceeds, one group gets frustrated and approaches you—they've come up with two solutions but can't agree on which one is correct. You see that one solution is right, while the other is not.
  b.) Describe both what would you do and what you would expect to happen as a result.
  c.) If the approach you described above in (b) didn't produce the result(s) you anticipated by the end of that class session, what would you do in the next class session?

---

*Figure 3.* Example scenario-based item on the FASCI

Most of the instruments reviewed which seek to measure some aspect of STEM teacher knowledge do so in a direct fashion, though observations of practice and follow-up interviews. An exception is the Mathematical Knowledge for Teaching (MKT) Instrument (?) (Hill, Schilling, & Ball, 2004), which includes multiple choice items t. The items on the MKT have a structure that presents hypothetical situations similar to the FASCI scenarios, but the MKT is much more focused on math teacher subject matter knowledge, rather than STEM strategic knowledge. Therefore the scenario-based, constructed-response structure of the FASCI items is somewhat unique.

This structure is not without precedence, however. In many studies on teachers' PCK, "critical events" are used as a prompt around which to discuss teacher actions (Hashweh, 1987; Shulman, 1986). Also, in the Mosaic II project, researchers at RAND developed what they call "vignette-based surveys" to measure science and math teachers' reform-oriented practices.  More specifically, these vignette-based surveys, or scenarios, were designed to measure a teacher's "intent to engage in reformed approaches" (Le et al., 2004). Separate scenarios were created for math and for science, and involve content specific to each domain. The scenarios are very specific for each grade level, subject, and the curriculum being taught by the teachers at each site.  Responses are not open-ended; rather the teacher rates the likelihood (on a scale of 1-4) that they would engage is a particular practice, given the scenario. For example, after being presented with specific information about a math problem and how a particular group has gone about approaching the problem, the respondent is asked how likely they are to "ask the class if they can think of another way to solve the problem" (on a scale of 1-4, where 1 = very unlikely and 4 = very likely). Between five and eight different rating questions are presented to the respondent in each scenario. Compared to these items, the scenario-based items on the FASCI are unique in the sense that they are not specific to a particular grade level, topic, and curriculum, and they require the respondent to construct an open-ended response.

**Validity Argument**

In developing a validity argument for the FASCI, it is important to first make explicit the proposed instrument use: to evaluate the effects of a teacher education program on novice STEM teachers' strategic knowledge (SK). This proposed use involves making both norm-referenced decisions (e.g., did a particular novice STEM teacher achieve a higher SK score than another novice STEM teacher?), and criterion-referenced decisions (e.g., did a particular novice STEM teacher achieve a certain level on the SK construct?). Score reliability is central to both of these situations, as precision in measurement is necessary in order to make such decisions. Given the structure of the FASCI,  a necessary proposition that is central to this validity argument is that SK can be measured reliably with a scenario-based survey. It is that proposition which is the focus of this paper.

Score reliability is often evaluated with Cronbach's alpha, which is a characterization of *relative* measurement error, and is of interest when making norm-referenced decisions such as that in the first situation posed above. Cronbach's alpha considers only one composite source of measurement error and does not take into account another potentially important source of error in the present study: that due to raters. Therefore it likely overestimates score reliability. But the classical test theory conception (represented by Cronbach's alpha) is not the only way to think about score reliability. By taking into account other *facets of measurement* such as the raters, and interactions between these facets, score reliability can be investigated more deeply. When considering the items, raters, and their interactions as potential sources of error, one can examine not only the reliability for making *relative* (norm-referenced) decisions, but also that for making *absolute* (or criterion-referenced) decisions.

In this paper, I will discuss the proposition that SK can be measured reliably with a scenario-based survey by investigating reliability in three ways: 1) rater agreement in scoring, 2) the classical test theory conception of score reliability, and 3) score reliability conceptualized within a Generalizability Theory (G Theory) (Brennan, 1992) framework. I will conclude by discussing the findings of the more rigorous G Theory reliability analysis and the implications of this work for other instrument development and validation efforts.

**Reliability in terms of Rater Agreement**

One approach that instrument developers often take to examining score reliability is to assess the agreement of raters in scoring open-ended responses or observable actions. For this work, three practicing teachers who were thought to be high on the SK construct were recruited and trained to rate FASCI responses.

During training, the raters added a level to the SCI scoring guide and worked to further clarify and define some of the scoring guide language. These raters talked about how they would need to come to agreement in scoring without the researcher pushing them to do so. Further, they saw the need for better definition on the scoring guides so that they could agree on scores in training, and made the necessary changes. The resulting FA and SCI scoring guides are shown in Figures 4 and 5 respectively, with the new or clarified language *italicized*. Very little change was made to the FA scoring guide—most of the work on this dimension was in having a discussion about what constituted a new strategy and an appropriate contextual factor, and in further defining some of the wording on the FA construct map. Level two on the SCI scoring guide was added to accommodate the new idea that a high-level response should include some *rationale* for why the situation was seen as an opportunity of interactive teaching and learning.

| Level | Modification of teaching approach | Discussion of *contextual* factors that bear on the modification of the teaching approach |
|---|---|---|
| 2 | YES | YES |
| 1 | YES | NO |
| 0 | NO | NO |

*Figure 4.* FA scoring guide

| Level | Discussion of interactive teaching | *Discussion of a rationale for why they see this as an interactive situation* |
|:-:|:-:|:-:|
| 2 | YES | YES |
| 1 | YES | *NO* |
| 0 | NO | *NO* |

*Figure 5.* SCI scoring guide

On the last independent scoring task in training, rater agreement was quite good on both the FA and SCI dimensions. Overall rater agreement for the full response set (60 responses) is shown in Tables 1 and 2 for the FA and SCI dimensions respectively. Rater agreement by individual item can be seen in Appendix B. On the FA dimension, agreement between pairs of raters ranged from 80%-90%. Cohen's kappa is also given for each pair of raters. This statistic is a bit more critical than percent agreement, in that it takes into account that rater agreement could have occurred by chance. On the FA dimension, kappa between pairs of raters ranged from 0.63 to 0.82. For the SCI dimension, agreement was not as high as that on the FA dimension. Percent agreement between pairs of raters ranged from 76%-88% and kappa ranged from 0.40 to 0.57.

Table 1.
*Overall FA Rater Agreement*

| Rater Combination | Percent Agreement | Cohen's Kappa |
|:-:|:-:|:-:|
| r1-r2 | 83% | .68 |
| r1-r3 | 80% | .63 |
| r2-r3 | 91% | .82 |

Table 2.
*Overall SCI Rater Agreement*

| Rater Combination | Percent Agreement | Cohen's Kappa |
|:-:|:-:|:-:|
| r1-r2 | 83% | .52 |
| r1-r3 | 76% | .40 |
| r2-r3 | 88% | .57 |

A first step towards a deeper examination of score reliability comes from computing Cronbach's alpha for the resulting scores.

**Reliability in terms of a Single Composite Source of Error**
Cronbach's alpha was calculated for each dimension of the FASCI. In using Cronbach's alpha to estimate score reliability, any measurement error cannot be attributed to scoring by the raters. This is because in classical test theory, the only source of measurement error is, in theory, due to the differences in scores that would be observed if the same respondent were to be surveyed over and over again repeatedly. There is no specific "rater component" of measurement error.

In classical test theory the value of the reliability coefficient (the ratio of true score variance to observed score variance) will be higher if the variability of the error component of observed scores is lower than the variability of the true score component (see equation 1). In this equation, $\sigma_\tau^2$ represents the variance in true score for an individual on one dimension of the FASCI, $\sigma_X^2$ represents the variance in observed score on that dimension, and $\sigma_e^2$ is the variance in the measurement error. This relationship shows that a decrease in error variance ($\sigma_e^2$) will increase score reliability.

$$reliability = \frac{\sigma_\tau^2}{\sigma_x^2} = \frac{\sigma_\tau^2}{\sigma_\tau^2 + \sigma_e^2} \qquad (1)$$

Score reliability resulting from FASCI responses in the present study was estimated and compared to results from previous pilot testing of the FASCI. This was done in order to see if score reliabilities resulting from this study are consistent with those from previous pilot tests, one of which had a significantly larger sample size from a similar population. Score reliability, sample size, and variability in observed scores (expressed in terms of standard deviations) from this study and from previous pilot tests are shown in Table 3. One can see that score reliability for each dimension is within the range of those found in earlier pilot testing of the FASCI. FA score reliability for this study was 0.62, and in the two previous pilot tests it was found to be 0.69 and 0.43. SCI score reliability was 0.51, and in the two previous pilot tests it was found to be 0.42 and 0.46. The SD of observed FA and SCI scores were lower than those observed in previous pilot testing. This is likely due to the fact the scores in this study are based on three raters, while in previous pilot studies scores were based on only one rater (with a 20% score behind check by a second trained rater). By averaging a respondent's score over three raters, there are likely to be fewer outliers (extremely high or low scores) that result from particularly harsh or easy scoring. This serves to decrease the spread of the observed scores. Note that there is a significant amount of missing data, which has an effect on score reliability.

Table 3.
*Reliability estimates (Cronbach's Alpha) and overall percentage of missing data for each dimension, this study and previous pilot testing*

|  | Sample size | FA alpha | SD of Observed FA score | % FA missing | SCI alpha | SD of Observed SCI score | % SCI missing |
|---|---|---|---|---|---|---|---|
| Present study | 60 | 0.62 | 0.37 | 21.7 | 0.51 | 0.36 | 13.3 |
| ¥FASCI Pilot test 1 | 63 | 0.69 | 2.05 | 6.5 | 0.42 | 1.13 | 5.5 |
| ¥FASCI Pilot test 2 | 96 | 0.43 | 1.73 | 2.4 | 0.46 | 1.24 | 1.7 |

¥Results from these pilot tests of the FASCI are reported in Author (2009).

According to Traub (Traub, 1994), reliability estimates of 0.80 or higher are routinely achieved for objectively scored tests, while those for subjectively scored tests (such as the FASCI) can be lower. This target value of 0.80 is similar to that achieved in a vignette-based surveys of mathematics teacher practice (Le et al., 2004), where estimates of internal consistency were observed to be between 0.80 and 0.86. However, as discussed above that survey was not open-ended, rather the respondents ranked the likelihood that they would engage in a particular behavior. Therefore it is reasonable to think that the lower end of that range (0.80) is a good target for an open-ended scenario-based instrument such as the FASCI. Given this benchmark, the score reliability observed for the FA and SCI dimensions is not very high, indicating a lack of internal consistency in this measure of SK.

As mentioned above, these estimates do not take into account any potential error due to the raters and therefore likely *overestimate* score reliability. Because of the previously observed low rater agreement (during instrument development), it is reasonable to think that there would be some error due to the raters. What is needed is a framework for decomposing the composite error variance of observed scores into separate components.

**Reliability in terms of Facets of Measurement**

In this deeper investigation of score reliability, we are interested in: 1) assessing how much of the variability in observed scores can be attributed to the items, raters, and interactions between respondents and these facets, and 2) how these sources of "error" variance could be reduced if the number of items and/or raters were to be increased. The first of these points is addressed by estimating error variances based on the observed scores, and will be discussed directly below. The second of these points involves estimating how score reliability and the standard error of measurement would change if the number of items and/or raters associated with FASCI scoring was changed.

These analyses are accomplished by applying Generalizability Theory (Brennan, 1992). In a G Theory analysis, multiple sources of error variance for single observations (e.g., a person's score on a single item scored by a single rater) can be estimated. This is done as a part of the generalizability study (or "G Study"). In the subsequent decision study (or "D Study") one then investigates the variance estimates associated with *mean* scores from some sample (e.g., of persons across a set of items and raters). Mean scores are the focus of the D Study as it is these scores that are the basis for any decisions made about a person. The D Study therefore helps to inform decisions about the optimal configurations for maximizing score reliability related to the proposed score interpretation and instrument use.

Sources of possible error variance are referred to as *facets of measurement* in G Theory. It is assumed that measurements are made from a *universe of admissible observations*, which are observations that are seen as interchangeable. Each facet is considered to have its own universe. For example, in this study the facets of interest are items and raters. Each of these is conceptualized as being from an infinite universe of admissible items or raters. In theory then, there are an infinite number of possible item scenarios from which those on the survey are drawn from, and an infinite number of potential raters of responses to those items. While perhaps not infinite, it is plausible to think that there are a very large number of potential teaching scenarios which could be included in a FASCI item, given the complexity of teaching. It is also plausible to think that there exists a very large pool of potential raters of FASCI responses. Conceiving of observed scores as being a sample of respondent performance drawn from a universe of potential item scenarios and rated by a universe of potential raters allows one to think of measurement error as being due to sampling variability. This is a key aspect of the G Theory approach. Accordingly, we would expect sampling variability to decrease as sample size increases. However, if all of the items in the universe are of similar quality, then scores resulting from a sample of items drawn from that universe will not differ greatly from scores resulting from any other sample drawn from the same universe. The same would hold true for raters. Ideally, we would want the bulk of observed score variability to be attributable to the *object of measurement* (in this case the persons, denoted by "*p*"). The facets of measurement that are not the object of measurement are therefore thought of as sources of error (in this case, items, denoted by *"i"* and raters, denoted by *"r"*), and accordingly we would want to minimize the role that the variance associated with each of those facets plays in FASCI score interpretations.

      G Theory represents a more nuanced way of thinking about score reliability than the classical test theory conception discussed above (Thompson, 2002). Rather than considering an observed score as being composed of a true score and an error score component (see equation 1), in a G Theory analysis one can consider the error attributable to multiple components (the facets). Because a G Theory analysis simultaneously estimates variance components for the object of measurement (e.g., persons) and multiple measurement facets (e.g., items and raters), the concept of observed score represented in equation 1 is somewhat different. Rather than thinking of true score and error score components of observed scores, we now think of *universe scores* and their variance components due to each facet of measurement.

      The item and rater facets are of interest because one would like to be able to generalize from one set of items or raters to a different (larger or smaller) set of each. Therefore in this study, the *universe of generalization* consists of different specified item/rater combinations. In making absolute decisions (i.e., whether or not a person achieves some specific SK score), each of these facets is a potential source of error in generalization. Items vary in difficulty and quality, so a respondent's score on one set of items may not be the same as that on another set. Raters vary in their judgments when scoring—some are harsher than others, for example. Each of these potential variances can be estimated in a G Theory analysis, and each are of particular interest when making absolute decisions. This is one of the proposed uses of the FASCI instrument as described above.

      As stated above, in this study items and raters (both conceptualized as being drawn from an infinite "universe" of each) are potential sources of error variability in *universe scores*. This study therefore has two facets of measurement. Further, the interactions between these facets and the object of measurement (persons), and the facets and each other, are other potential sources of variability. Because we would accept within the universe of admissible observations the response of any person (*p*) to any item *(i)* scored by any rater *(r)*, then the G Study design is said to be *fully crossed*. This is represented by the notation *p x i x r* (read "p cross i cross r", or "p by i by r"). This

fully crossed design, each of the variance components, and their interactions is represented by the Venn diagram in Figure 6.

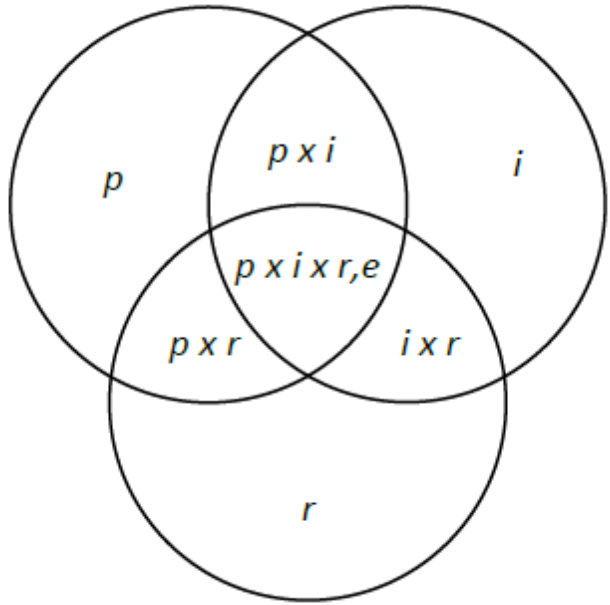

*Figure 6.* Venn diagram representing variance components in the *p x i x r* design G Study

Once each of these variance components is estimated in the G Study, two types of error variance can be calculated in the G and D Studies: *absolute error variance* and *relative error variance*. Absolute error variance is denoted by uppercase delta (Δ), and is the sum of all variance components which involve the *facets* of measurement $(\hat{\sigma}_I^2, \hat{\sigma}_R^2, \hat{\sigma}_{pI}^2, \hat{\sigma}_{pR}^2, \hat{\sigma}_{iR}^2, \hat{\sigma}_{pIR,e}^2)^2$. The square root of the absolute error variance is the absolute standard error of measurement (SEM Δ). Absolute error variance is of interest when making criterion-referenced decisions, such as whether or not a person achieves a pre-determined SK score that is representative of some level on the SK construct. Relative error variance is denoted by lower case delta (δ), and is the sum of all variance components which involve the object of measurement $(\hat{\sigma}_{pI}^2, \hat{\sigma}_{pR}^2, \hat{\sigma}_{pIR,e}^2)$. Again, the square root of this error variance is the SEM (δ). Relative error variance is of interest when comparative decisions are being made, such as distinguishing one respondent from another. Each of these error variances, errors, and SEMs are calculated for a particular D Study, and can therefore be compared across different D Studies which specify different numbers of items and/or raters.

These errors are also summarized in two different reliability coefficients: the generalizability coefficient ($\rho^2$) and the dependability coefficient (Φ) (Brennan 1992). $\rho^2$

---

[2] The "hat" (^) over each variance component in the D Study indicates that they are new estimates of the parameters obtained in the G Study. Also, note the capitalized subscripts for items and raters (*I* and *R*) indicating that these are variance components associated with mean scores rather than individual observations (as in the G study).

is analogous to a reliability coefficient in classical test theory, which is a characterization of *relative* error. This coefficient is of interest when making comparative decisions. The generalizability coefficient takes into account the error associated with all components that involve the object of measurement (see equation 2). In this equation, $\sigma_\tau^2$ represents the universe score variance. In contrast to the conceptualization of score reliability presented in equation 1 which had only a single term in the denominator denoting error variance, the equation for $\rho^2$ has three separate terms (in the context of this study) in the denominator which represent sources of error. Therefore, for a single facet G Study $\rho^2$ would be equivalent to Cronbach's alpha. But for a multiple facet design (as in this study), they are not comparable. Classical test theory partitions variance into only two sources (true score and error score), while G Theory partitions variance into multiple sources (universe score and each variance component involving the object of measurement) (Shavelson & Webb, 1991; Thompson, 2002).

$$\rho^2 = \frac{\sigma_\tau^2}{\sigma_\tau^2 + \hat{\sigma}_{pI}^2 + \hat{\sigma}_{pR}^2 + \hat{\sigma}_{pIR,e}^2} \qquad (2)$$

G Theory analyses also provide another estimate of score reliability, the dependability coefficient (Φ). This coefficient is of interest when making absolute decisions. The dependability coefficient Φ is therefore a characterization of *absolute* error, and as such takes into account the error associated with each facet of measurement (equation 3). Because Φ takes into account the same variance components in the equation for $\rho^2$ plus the variance components that are associated with the facets and their interactions, the value for Φ is generally lower than that for $\rho^2$. In this sense, it is a more critical estimate of score reliability, as one would expect for making absolute decisions rather than relative ones.

$$\Phi = \frac{\sigma_\tau^2}{\sigma_\tau^2 + \hat{\sigma}_I^2 + \hat{\sigma}_R^2 + \hat{\sigma}_{pI}^2 + \hat{\sigma}_{pR}^2 + \hat{\sigma}_{IR}^2 + \hat{\sigma}_{pIR,e}^2} \qquad (3)$$

All G Theory analyses were completed using the software GENOVA (Crick & Brennan, 2001). All item scores in the form of .dat files, as well as GENOVA model specification and D study design files from this study are available for reader examination and downloading/use in an online repository[3].

**G Study Estimates of the Variance Components.**
Figure 7 shows the percentage of observed score variance attributable to the object of measurement and to each facet for the FA dimension of SK as estimated in the G study. Table 4 shows the specific variance components estimates and the percentage that each component contributes to the overall variance for both the FA and SCI dimensions. While relatively little variance is attributable to the items themselves, a very large portion of the variance in observed scores (almost 60%) is in the person by item (*p x i*) interaction. This indicates that a respondent's FA scores across the different raters depend heavily on the particular item being sampled from the universe of items (i.e., the teaching scenario to which they are responding). This finding is consistent with related analyses of students' responses to science performance assessments (Ruiz-Primo & Shavelson, 1996; Shavelson, Baxter, & Gao, 1993). Very little variance in observed FA scores is attributable to the raters, and almost none to the *p x r* and *i x r* interaction terms. This indicates that the raters were fairly consistent in their ratings across persons and items, an observation that is supported by the relatively strong rater agreement in scoring.

---
[3] Before publication, a link to a github repository which contains these files will be made available. It is not included now in order to protect the author's identity.

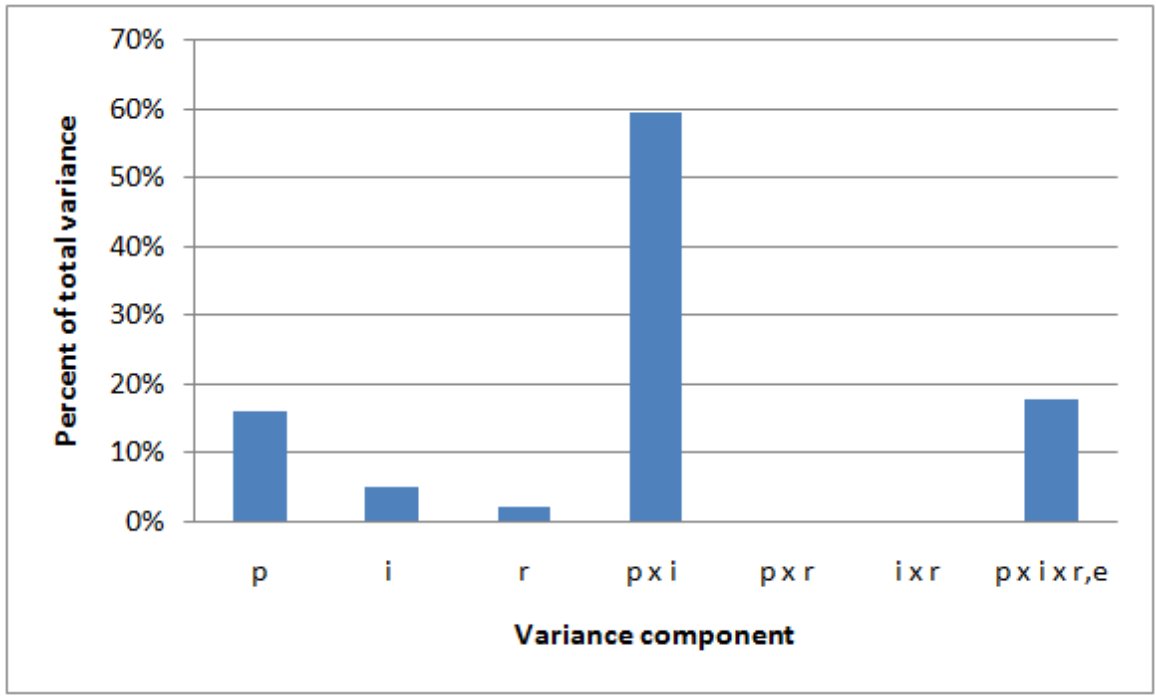

*Figure 7.* Sources of variability in FA scores

Table 4.
*Variance estimates and percentage of total variance, FA and SCI dimensions*

| Component | FA Variance | FA % of total | SCI Variance | SCI % of total |
|---|---|---|---|---|
| *p* | 0.064 | 15.9% | 0.053 | 7.1% |
| *i* | 0.02 | 5.0% | 0.315 | 42.5% |
| *r* | 0.008 | 2.0% | 0.014 | 1.9% |
| *p x i* | 0.24 | 59.6% | 0.234 | 31.5% |
| *p x r* | 0 | 0.0% | 0.006 | 0.8% |
| *i x r* | 0 | 0.0% | 0.002 | 0.3% |
| *p x i x r,e* | 0.071 | 17.6% | 0.118 | 15.9% |

Figure 8 shows the percentage of observed score variance attributable to the object of measurement and to each facet for the SCI dimension of SK. In contrast to the findings for the observed FA scores, a large part of the variance in observed SCI scores (42%) can be attributed to the items themselves. In other words, the mean score for one randomly selected item (across all persons and raters) is expected to be quite different from the mean score for all items in the universe (across all persons and raters). Conceived of in this way, this error due to items can be thought of as *sampling error*. A particular item (a teaching scenario) which is sampled from the universe of potential items and included on the survey is likely to be much more difficult or easy for respondents than the average across all of the potential items. Consistent with the variance in observed FA score discussed above, there is also a large amount of variance in observed SCI scores (32%) that can be attributed to the person by item (*p x i*) interaction. This means that a respondent's SCI score across all raters depends on the particular item to which they are responding. Finally, a relatively small amount of the variance in observed SCI scores is attributable to the raters (~2%). Again, this is consistent with the rater agreement on the SCI dimension which was quite good, but not as good as that on the FA dimension.

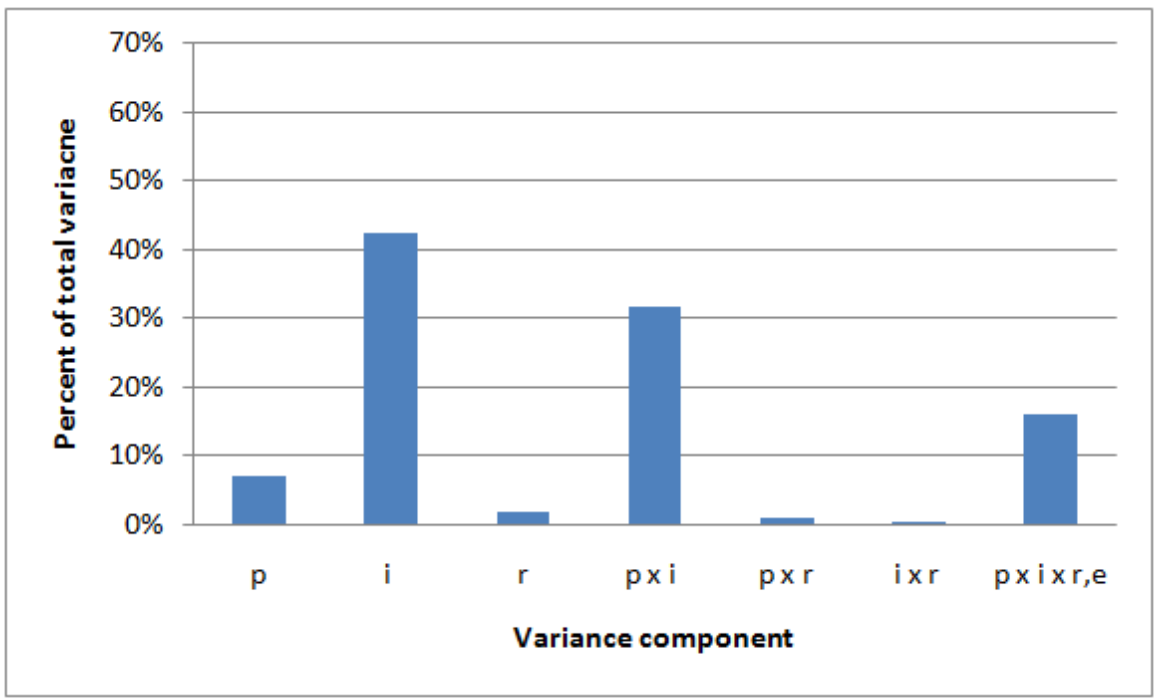

*Figure 8.* Sources of variability in SCI scores

The central point to highlight from both of these analyses is the importance of the items chosen for use on the instrument. Respondent performance can vary quite a bit depending on the particular item or set of items to which they respond, and the mean score of all persons across any one item is likely to be very different than the mean score of all persons across a larger set of items. It is also important to emphasize that the items in this case are the teaching scenarios, not the constant item prompts to which individuals respond within each scenario. Again, these findings are consistent with previous work on science performance assessments, where the researchers found that item-sampling variability was a considerable source of measurement error (Ruiz-Primo & Shavelson, 1996; Shavelson et al., 1993). The implication therefore is that a large number of items are needed on the instrument in order to obtain a reliable measure of respondent performance.

**Minimizing the Role of Error Variance in Score Interpretations.**

The purpose of the G study was to estimate variance components that were associated with a single observation from the *universe of admissible observations* (e.g., a person's score on one item rated by one rater). These estimates were discussed above in Figures 4 and 5. In the subsequent D studies, the context now shifts to the *universe of generalization*: a specified universe of measurement procedures to which one wants to generalize based on the results of the study. The universe of generalization in this case is specified as different combinations of items and/or raters (beyond the observed five items and three raters). In the D study, variance estimates are generated for *mean* scores of persons, in this case across items and raters. These D study variance components are estimated using the G study variance estimates. For example, the G study variance estimate for items on the SCI dimension was 0.315. This estimate is based on a single person-item-rater observation. The associated D study variance component for a five item specification on the SCI dimension would be 0.315/5 = 0.063. This estimate is for mean scores across items. The expected value of these mean scores for a specified measurement procedure is a person's *universe score*. These D study variance components are the basis for computing measurement error and score reliability that would theoretically result from these different measurement specifications (i.e., item/rater combinations). For both the FA and SCI dimensions of strategic knowledge, D studies were specified for all combinations of five to ten items and one to five raters, and generated both relative and absolute error variances and reliability estimates for each D study.

The upper panel of Figure 9 shows how the generalizability coefficient (from here on referred to as "relative reliability") varies as a function of number of items and raters for the FA dimension. For the baseline five item/three rater combination, relative reliability was estimated to be 0.54, which is lower than the value for Cronbach's alpha discussed above (0.62 for this study). This value is smaller than that for Cronbach's alpha because the relative reliability estimate now takes into account facet interactions (*p x i*, *p x r*, and *p x i x r,e*), which are in the denominator of the reliability coefficient calculation, and therefore decreases the value of the estimate. These two facet interactions accounted for an estimated 77% of the variance in observed scores based on the G Study (see Table 4). In this context, by not taking the rater facet interactions (*p x r* and *p x i x r,e*) into account, we would be over-estimating score reliability. This is precisely why the value for Cronbach's alpha is higher.

Not surprisingly, the results from this analysis indicate that the best way to improve FA score reliability comes from adding items to the instrument, rather than from adding raters. Beyond two raters, there is not much increase in reliability for a given number of items. This is consistent with Figure 4 above, which shows the largest amount of variability in observed scores was in the person by item interaction, indicating that a respondent's FA score across raters depends on the particular item being sampled. In order to ameliorate that effect, a larger number of items would be needed on the instrument. Increasing the number of items "sampled" from the universe of items would decrease the standard error of measurement for FA scores. This will be illustrated further below when specifically discussing the standard errors of measurement. Specifically, by doubling the number of items to ten and keeping the number of raters at three, relative reliability could be increased to about 0.71. A similar increase could also be realized by doubling the number of items and having just two raters rather than three. Because relatively little variability in observed scores was attributable to the raters, adding raters (i.e., increasing the size of the rater sample) would not decrease the overall role that error variance plays. In other words, averaging over more raters would not produce the same effect as averaging over more items because of the low variability attributable to raters compared to that of the items.

The dependability coefficient (from here on referred to as "absolute reliability") is of interest when making absolute or criterion-referenced decisions, which is consistent with one of the proposed uses of the FASCI instrument. In the D Studies, estimates of absolute reliability are generated for different combinations of items and raters. For five items and three raters, absolute reliability for FA scores is 0.52. The lower panel of Figure 9 shows the estimates of absolute reliability as a function of number of items for different rater combinations on the FA dimension. Comparing this to upper panel of Figure 9 (relative reliability estimates) shows that these estimates for absolute reliability are similar, though a bit lower. They are slightly lower due to the fact that when making absolute decisions, more facet conditions are taken into account: those involving the facets and their interactions as well as those involving the object of measurement (see equations 2 and 3 above). Because of this, measurement error is increased and the related reliability estimates are lower (i.e., more critical) than those for making relative decisions.

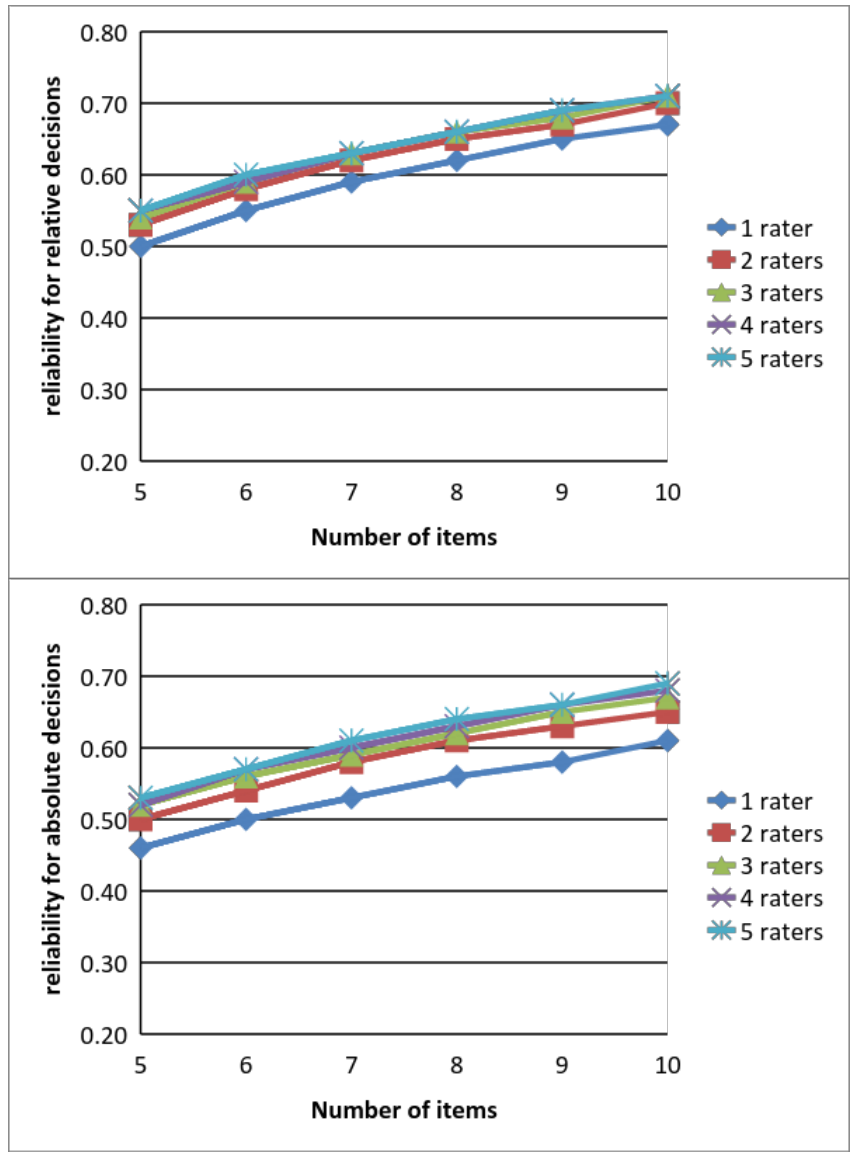

*Figure 9.* Relative and absolute reliability estimates as a function of number of items and number of raters, FA dimension

Figure 10 shows the same plots for the SCI dimension: relative reliability estimates as a function of number of items and raters in the upper panel and absolute reliability estimates in the lower panel. Because there was more variance attributable to raters and rater interactions on this dimension (consistent with the slightly lower rater agreement on SCI compared to FA), a larger increase in relative reliability (compared to the FA dimension) is realized when scores are averaged over additional raters, though once again there are clearly diminishing returns. As with the FA dimension, the largest increases in SCI relative reliability can be realized by adding items to the instrument. For five items and three raters, relative reliability is estimated to be 0.48, which is lower than the value for Cronbach's alpha for these same scores (0.51). By increasing the number of items to ten and keeping the number of raters at three, relative reliability could theoretically be increased to about 0.64.

The lower panel of Figure 10 shows the absolute reliability estimates for the SCI dimension. Comparing this to upper of Figure 10 again shows that the estimates of absolute reliability are lower than those for relative reliability. For five items and three raters, absolute reliability for SCI scores is 0.30. And comparing the lower panel of Figure 10 to the lower panel of Figure 9 shows that the absolute reliability estimates for SCI are *much lower* than those observed for the FA dimension. This is because of the larger item and rater variance component for SCI, and indicates that it would be very difficult to make any absolute decisions on the basis of SCI scores, a finding that is further supported in the discussion of standard error of measurement (below).

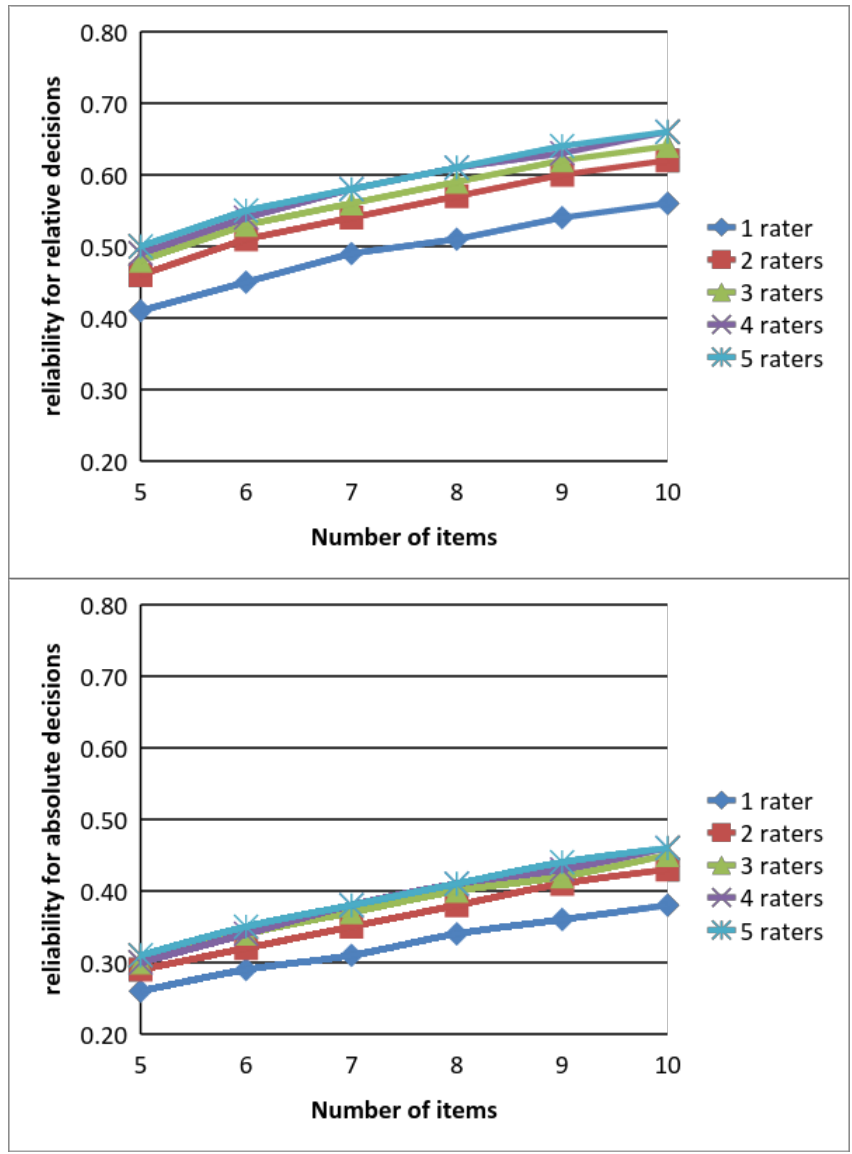

*Figure 10.* Relative and absolute reliability estimates as a function of number of items and number of raters, SCI dimension

It should be noted that in these analyses, it is assumed that all items in the universe of generalization have been well-specified and as such are of similar quality. Therefore if a poorly discriminating item has been developed and is one of the items sampled from the universe when increasing the number of items, then the expected gain in reliability will not be realized[4]. In other words, adding poor quality items will not help. Increasing the number of items and/or raters on the instrument also comes at a cost of respondent time and/or rater time and money to pay raters. Such changes would need to be evaluated relative to available resources, proposed uses of the instrument and scores, and other factors. For example, the current respondent time burden for a five item set is about 35 minutes. Several respondents noted that they became frustrated with the repetitive item prompts from one scenario to the next (e.g., *"I didn't know in the last question and I don't know now"* ID 1977205, in response to item 5). Based on this observation, it is reasonable to believe that respondent fatigue could play a role when increasing the number of items. If this were the case, then doubling the number of items is not likely to result in these theoretical increases in reliability.

In order to examine the impact of increasing the number of items on measurement error, one can plot the standard errors of measurement for absolute decisions (SEM Δ) and that for relative decisions (SEM δ) as a function of number of items. Figure 11 shows this plot for the FA dimension, and Figure 12 for the SCI dimension. In both plots, the number of raters is held constant at three (though these values are very similar for two raters).

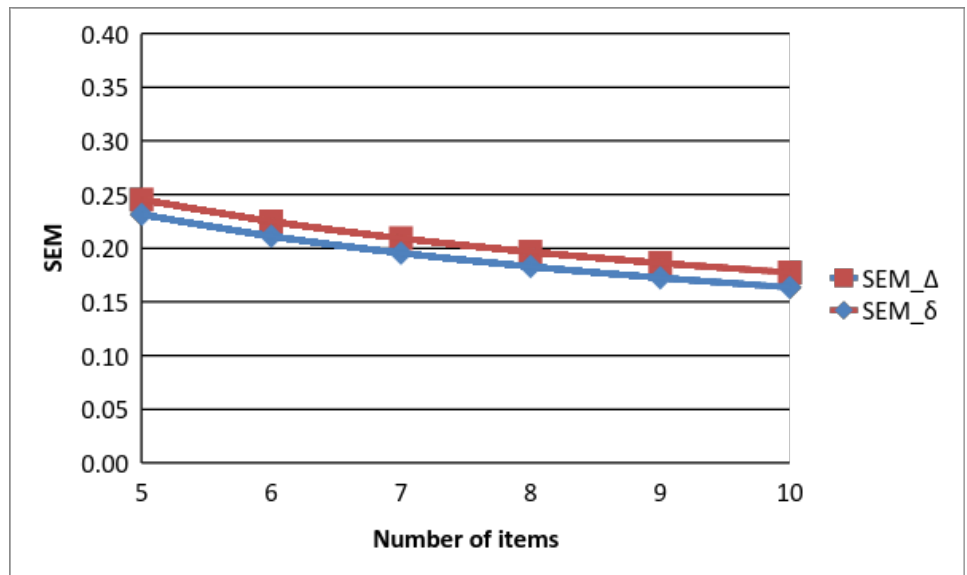

*Figure 11.* Plot of standard errors of measurement for absolute (Δ) and relative (δ) decisions, FA dimension

---

[4] Item discrimination was estimated by calculating the Pearson correlation between the average item score for an individual (across all raters) and their average total score (Crocker & Algina, 1986). SCI item discriminations were 0.63, 0.61, 0.42, 0.66, and 0.65. FA item discriminations were 0.74, 0.59, 0.63, 0.49, 0.65.

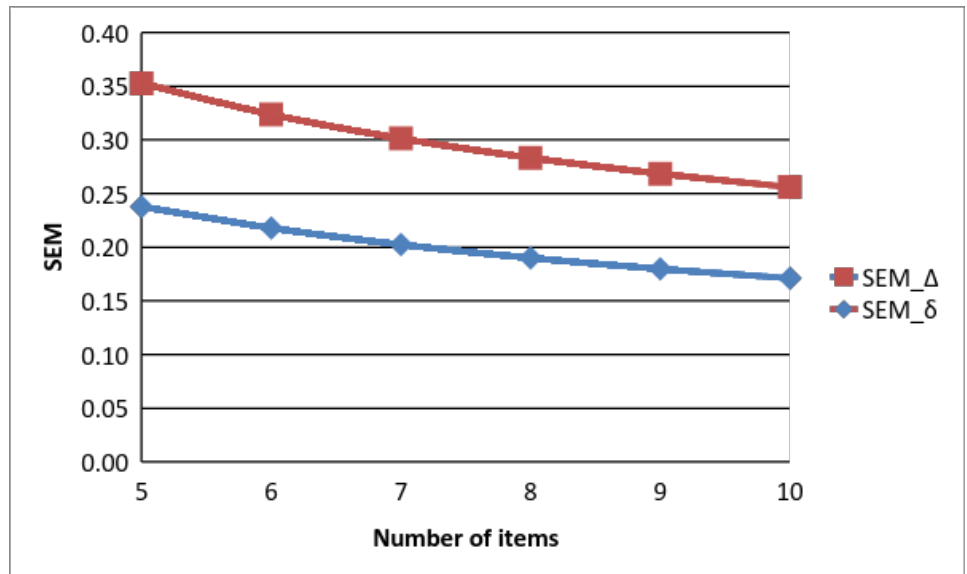

*Figure 12.* Plot of standard errors of measurement for absolute (Δ) and relative (δ) decisions, SCI dimension

In each of these plots, one can see that increasing the number of items decreases the standard error of measurement for both absolute (criterion-referenced) and relative (norm-referenced) decisions. This is consistent with the sampling framework conception: as sample size increases, SEM decreases. Because the SEM is inversely related to the square root of the sample size, increasing the sample size by a factor of four will decrease the SEM by a factor of two. Therefore in the present context, we would expect doubling the number of items in the sample of items to decrease the SEM by a factor of $\sqrt{2} = $ *1.41*.

For relative decisions, the standard error of measurement (SEM δ) as a function of number of items is very similar for FA and SCI. This is because for both dimensions, a large amount of the variance was observed to be in the person by item interaction, and very little was in the person by rater interaction. This statistic is based on the variance components which involve the object of measurement (persons) as discussed above.

The standard error of measurement for making absolute decisions (SEM Δ) is quite different between the FA and SCI dimensions. For the SCI dimension, there is a larger standard error of measurement (by about 30%) for making absolute decisions as compared to that for the FA dimension. This is due to the fact that this statistic is based on the variance components which involve the facets of measurement, and for the SCI dimension a large amount of the variance was observed to be in the items themselves (which was not the case for the FA dimension). Some SCI items were either very easy or very hard for respondents. Therefore on the SCI dimension it would be particularly difficult to make absolute (criterion-referenced) decisions about respondent knowledge with a small number of items, an observation that was made above when examining the absolute reliability for this dimension. For example, the SEM Δ for five items and three raters on the SCI dimension is about 0.35. This means that a 95% confidence interval around a respondent's SCI score would be roughly +/- 0.70. This is a very large range with respect to the observed range of SCI scores, which is from 0 to 1.67 (see Table 5).

Even if the number of items was doubled to ten, the SEM Δ is still about 0.25 (a decrease by a factor of $\sqrt{2}$), and the 95% confidence interval would be about +/- 0.50. Such broad confidence intervals would make it difficult to claim that the SCI score observed for a particular respondent had not occurred by chance. In other words, a very different SCI score would likely be observed for that same individual had they responded to a different set of SCI items.

Table 5.
*Range, mean, and SD of observed FA and SCI scores*

|         | FA   | SCI  |
|---------|------|------|
| minimum | 0    | 0    |
| maximum | 1.44 | 1.67 |
| mean    | 0.64 | 0.67 |
| SD      | 0.37 | 0.36 |

To further illustrate the magnitude of SEM Δ for SCI scores, consider Figure 13 which shows all 60 mean SCI scores plotted with error bars representing a 95% confidence interval of +/- 0.70 (based on the SEM Δ for five items and three raters of 0.35). If a score of 1 on SCI represents the middle level of that construct (the teacher views the situation as an opportunity for interactive teaching and learning, but does not articulate a rationale why it is viewed as such), could one say with confidence that someone with a mean SCI score of 1 was emblematic of that level on the construct? With a 95% confidence interval of +/-0.70 around that mean score of 1, one could not make such a claim. The 95% confidence interval around this score ranges from 0.30 to 1.70. However, for lower-stakes decisions, a more relaxed confidence interval (e.g., 68% corresponding to +/- 1 SEM) could be used which would make it easier to distinguish respondents.

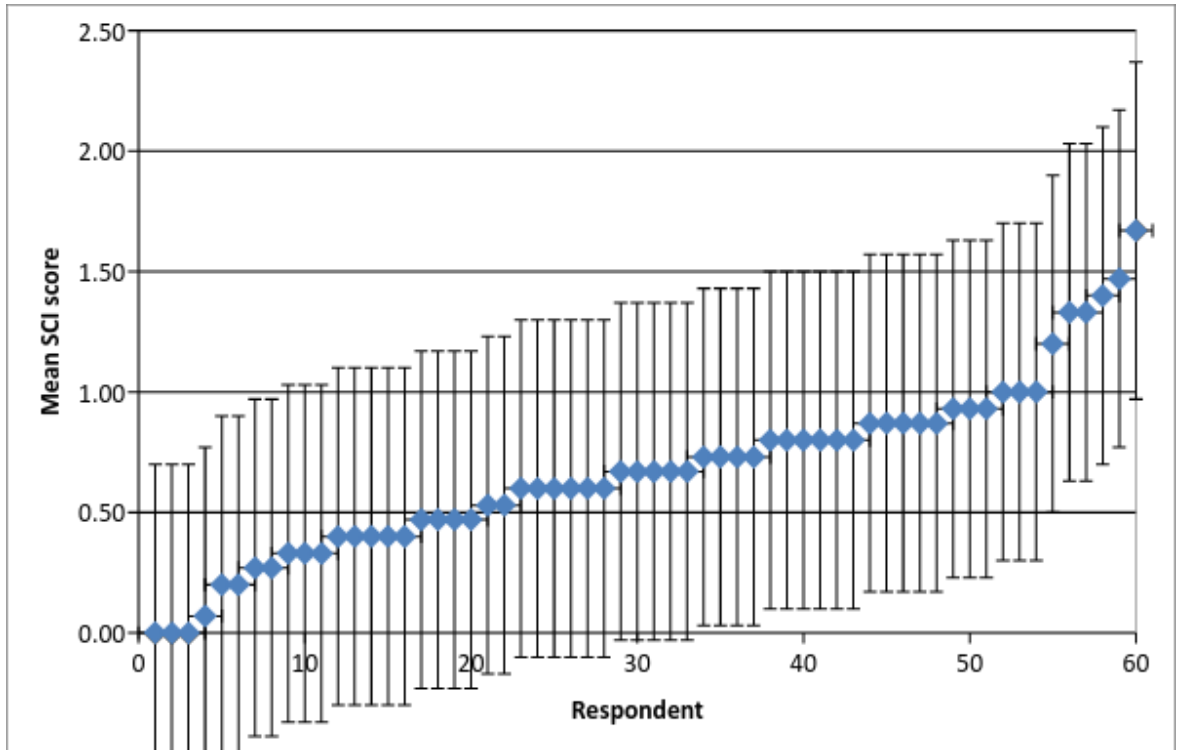

*Figure 13.* Mean SCI scores from all respondents with error bars representing a 95% confidence interval of +/-0.70 (SEM Δ = 0.35).

The SEM δ (for relative decisions) is very similar for both FA and SCI, as stated above. Each is about 0.23 for five items and three raters, meaning that a 95% confidence interval around a respondent's score would be +/- 0.46. As an example, consider the range of this 95% confidence interval for five items and three raters (0.92) around a particular observed SCI score. The range of observed SCI scores is 0 to 1.67 (Table 5). Therefore one can distinguish about two groups (0.92/1.67 = 0.55) on SCI with this confidence interval. By doubling the number of items to ten, SEM δ decreases to about 0.17 for each dimension, meaning that a 95% confidence interval now spans +/- 0.34. However, even if this decrease in SEM δ were realized, one could still not distinguish three groups on SCI (0.68/1.67 = 0.41). This lower SEM for relative decisions is because score precision does not need to be quite as high as that for making absolute decisions.

As a further example, consider the plot of all 60 mean SCI scores surrounded by a 95% confidence interval based on SEM δ = 0.23 (Figure 14). Many respondents in the top quartile of SCI scores (which ranges from 1.11 to 1.67) overlap with many respondents in the middle and bottom quartiles. In other words, there is not enough precision in measurement to distinguish each of these quartiles and therefore to make relative claims about the SCI scores of most respondents. This plot looks much the same for FA (since SEM δ is the same for that dimension; Figure 15) but relative comparisons are even more tenuous on that dimension due to the smaller range of observed scores (0 to 1.44). On the FA dimension, the bottom end of the top quartile

(which ranges from 0.96 to 1.44) overlaps with many scores in the bottom quartile (which ranges from 0 to 0.48).

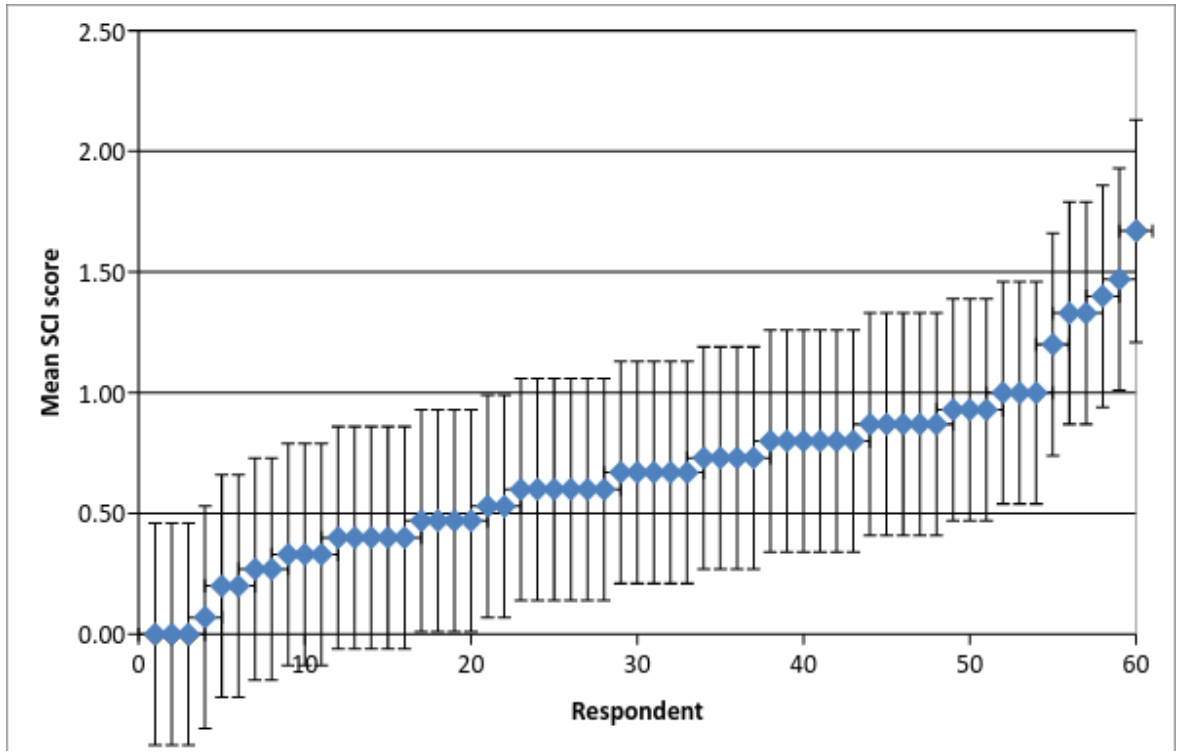

*Figure 14.* Mean SCI scores from all respondents with error bars representing a 95% confidence interval of +/-0.46 (SEM ð = 0.23).

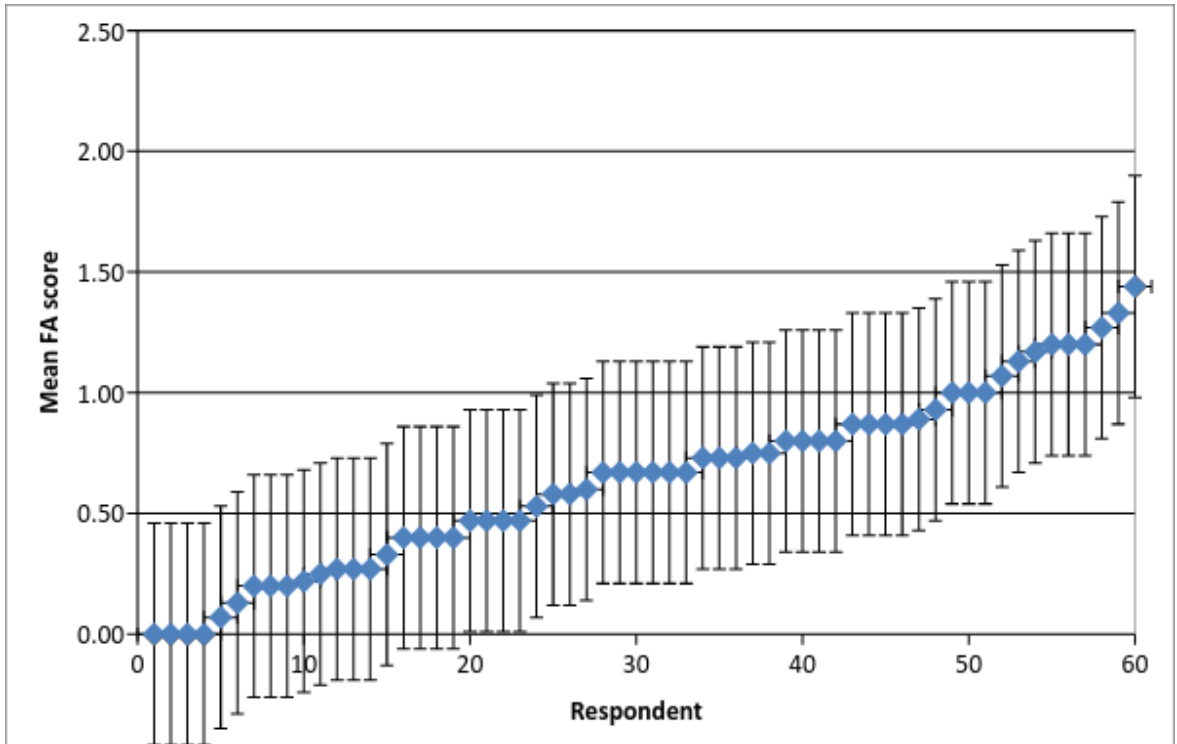

*Figure 15.* Mean FA scores from all respondents with error bars representing a 95% confidence interval of +/-0.46 (SEM δ = 0.23).

**Discussion**

    Score reliability for open-ended responses or observable actions is often thought of in terms of the agreement between raters. In some instrument development projects, high enough rater agreement might constitute "reliability." The RTOP (Sawada et al., 2002) is an example of this. Scoring agreement between the raters in this study was observed to be fairly good, but this agreement provides only a very coarse picture of score reliability.

    As a first step into a more critical examination, score reliability was conceptualized in terms of the ratio of true score variance to observed score variance. The starting point for this examination was Cronbach's alpha, which only considers one source of measurement error. Reliability as characterized by Cronbach's alpha was observed to be far below the target value of 0.80. A deeper examination of score reliability is realized by taking a G Theory approach, in which error variance can be decomposed into multiple sources. Further, using G Theory, score reliability for both relative and absolute decisions can be characterized.

    In the D studies, I was able to estimate how score reliability could potentially increase if different measurement procedures were specified (i.e., different combinations of items and/or raters). Based on the G study finding that much of the variance in observed scores was found in the items and person by item interaction, it is clear that the number of items included on the instrument is the driving force in reliability. This was found to be the case for both dimensions. Adding more raters will not increase reliability estimates appreciably, but adding items to the instrument will. If the number of items on the instrument was doubled (from five to ten), relative reliability

estimates increase to about 0.71 on the FA dimension and 0.64 on the SCI dimension. However, these values are still rather low compared to the target value of 0.80 presented above. Absolute reliability estimates were lower than the relative reliability estimates, because they include both the facets and facet interactions as potential sources of error. These findings raise a serious issue about the reliability (both relative and absolute) of SK scores from the FASCI. Even doubling the number of items would not yield high estimates of score reliability, assuming that these increases could be realized. As mentioned above, this increase in number of items comes at the cost of respondent time burden. Average time for respondents to complete the five-item version of the FASCI was around thirty five minutes. Requiring seventy minutes of respondent time would be asking too much of their participation, given that they will likely become fatigued and response quality will decrease.

In order to further illustrate these issues with this lack of precision in measurement, I estimated the standard errors of measurement for both relative and absolute decisions as a function of number of items. The SEM for relative decisions (SEM $\delta$) is similar for FA and SCI, and decreases as the number of items increases. For both dimensions, it would be difficult to make relative decisions (e.g., comparing the top and bottom quartiles) with the current number of items and raters. The SEM for absolute decisions (SEM $\Delta$) is much higher for the SCI dimension due to the fact that SCI items were either very easy or very hard for respondents (i.e., they were not equally discriminating). It would be very difficult to defend an absolute (criterion-referenced) decision for an individual on the SCI dimension. This finding is also reflected in the estimates of the absolute reliability for the SCI dimension.

**Conclusion**

Investigating score reliability is very important part of the measurement process, and it is central to the development of a validity argument. The degree of precision which is required of a measure depends on its proposed use, and it affected by many potential aspects of measurement. In this work, I have taken a multi-faceted approach to scrutinizing the reliability of an open-ended, scenarios-based measure of novice STEM teachers' strategic knowledge. Rather than relying on conventional measures and accepted values of interrater agreement or reliability coefficients which do not decompose sources of measurement error, I conducted a G Theory analysis and discussed the results. A particular affordance of this approach is being able to examine reliability for both absolute (e.g., criterion referenced) and relative (e.g., norm referenced) decisions. Further, conducting decision studies ("D Studies") allow one to model how score reliability for absolute and relative decisions might change by changing the number of measurement facets (in this case, items and raters).

Science education researchers who are developing and using measurement instruments need to take a serious look at the reliability of scores in order to establish validity for the measure. A deeper investigation into reliability can be insightful, and can better inform how one should proceed with further development and use of the measurement tool. Ultimately, the inferences that one can make based on a measure are only as strong as the validity argument for the measurement tool. And that validity argument depends at least in part on the argument for score reliability.

**Appendix A: FASCI items**

**For the questions and scenarios that follow, please assume that you are teaching a high school course in physics, chemistry, biology, Earth science or math to a class of 25-30 students.**

1. Students are working in groups of four to discuss a conceptual question you provided them at the beginning of class.
   a. How might this activity facilitate student learning?

As the activity proceeds, one group gets frustrated and approaches you—they've come up with two solutions but can't agree on which one is correct. You see that one solution is right, while the other is not.

   b. Describe both what would you do and what you would expect to happen as a result.
   c. If the approach you described above in (b) didn't produce the result(s) you anticipated by the end of that class session, what would you do in the next class session?

2. You are working out an example problem up on the board.
   a. How might this activity facilitate student learning?

You accidentally make a mistake in solving the problem but don't realize this until you get to the end of your solution and realize that the answer doesn't make sense. No one in the class has said anything, so you're not sure if they caught the mistake or not.

   b. Describe both what would you do and what you would expect to happen as a result.
   c. If the approach you described above in (b) didn't produce the result(s) you anticipated by the end of that class session, what would you do in the next class session?

3. You have just finished giving a presentation on a complicated topic.
    a. How might this activity facilitate student learning?

You notice that many of the students in the class have very confused expressions on their faces.

    b. Describe both what would you do and what you would expect to happen as a result.
    c. If the approach you described above in (b) didn't produce the result(s) you anticipated by the end of that class session, what would you do in the next class session?

4. You have given your students a quiz to assess their understanding of a difficult topic.
    a. How might this activity facilitate student learning?

Many of your students are discouraged after performing poorly on the quiz.

    b. Describe both what would you do and what you would expect to happen as a result.
    c. If the approach you described above in (b) didn't produce the result(s) you anticipated by the end of that class session, what would you do in the next class session?

5. In talking with one of your students you discover that they have a misconception about a central topic presented in that week's class. You attempt to address the misconception by having a one-on-one conversation with the student.
    a. How might this activity facilitate student learning?

Despite your conversation, the student maintains the same misconception.

    b. Describe both what would you do and what you would expect to happen as a result.
    c. If the approach you described above in (b) didn't produce the result(s) you anticipated by the end of that class session, what would you do in the next class session?

## Appendix B: Rater Agreement by item

FA Rater Agreement

*Table B1.*
Overall FA Rater Agreement

| Rater Combination | Percent Agreement | Kappa |
|---|---|---|
| r1-r2 | 83% | .68 |
| r1-r3 | 80% | .63 |
| r2-r3 | 91% | .82 |

*Table B2.*
Item 1 FA Rater Agreement

| Rater Combination | Percent Agreement | Kappa |
|---|---|---|
| r1-r2 | 82% | .69 |
| r1-r3 | 77% | .61 |
| r2-r3 | 92% | .85 |

*Table B3.*
Item 2 FA Rater Agreement

| Rater Combination | Percent Agreement | Kappa |
|---|---|---|
| r1-r2 | 88% | .78 |
| r1-r3 | 82% | .66 |
| r2-r3 | 87% | .73 |

*Table B4.*
Item 3 FA Rater Agreement

| Rater Combination | Percent Agreement | Kappa |
|---|---|---|
| r1-r2 | 80% | .64 |
| r1-r3 | 80% | .62 |
| r2-r3 | 92% | .83 |

*Table B5.*
Item 4 FA Rater Agreement

| Rater Combination | Percent Agreement | Kappa |
|---|---|---|
| r1-r2 | 82% | .62 |
| r1-r3 | 78% | .56 |
| r2-r3 | 93% | .85 |

*Table B6.*
Item 5 FA Rater Agreement

| Rater Combination | Percent Agreement | Kappa |
|---|---|---|
| r1-r2 | 83% | .65 |
| r1-r3 | 85% | .68 |
| r2-r3 | 93% | .84 |

SCI Rater Agreement

*Table B7.*
Overall SCI Rater Agreement

| Rater Combination | Percent Agreement | Kappa |
|---|---|---|
| r1-r2 | 83% | .52 |
| r1-r3 | 76% | .40 |
| r2-r3 | 88% | .57 |

*Table B8.*
Item 1 SCI Rater Agreement

| Rater Combination | Percent Agreement | Kappa |
|---|---|---|
| r1-r2 | 85% | .72 |
| r1-r3 | 78% | .62 |
| r2-r3 | 90% | .83 |

*Table B9.*
Item 2 SCI Rater Agreement

| Rater Combination | Percent Agreement | Kappa |
|---|---|---|
| r1-r2 | 82% | .34 |
| r1-r3 | 82% | .29 |
| r2-r3 | 92% | .54 |

*Table B10.*
Item 3 SCI Rater Agreement

| Rater Combination | Percent Agreement | Kappa |
|---|---|---|
| r1-r2 | 87% | .39 |
| r1-r3 | 87% | .42 |
| r2-r3 | 92% | .40 |

*Table B11.*
Item 4 SCI Rater Agreement

| Rater Combination | Percent Agreement | Kappa |
| --- | --- | --- |
| r1-r2 | 73% | .40 |
| r1-r3 | 63% | .21 |
| r2-r3 | 83% | .41 |

*Table B12.*
Item 5 SCI Rater Agreement

| Rater Combination | Percent Agreement | Kappa |
| --- | --- | --- |
| r1-r2 | 87% | .76 |
| r1-r3 | 68% | .47 |
| r2-r3 | 82% | .69 |